\documentclass[galaxies,article,accept,pdftex,moreauthors]{Definitions/mdpi} 
\firstpage{1} 
\pubvolume{1}
\issuenum{1}
\articlenumber{0}
\pubyear{2025}
\copyrightyear{2025}
\externaleditor{~ } 
\datereceived{14 August 2025 } 
\daterevised{8 September 2025 } 
\dateaccepted{9 September 2025 } 
\datepublished{ } 
\hreflink{https://doi.org/} 

\Title{The Murchison Widefield Array Enters Adolescence: A Personal Review of the Early Years of Operations}

\TitleCitation{The Murchison Widefield Array Enters Adolescence: A Personal Review of the Early Years of Operations}

\Author{Steven J. Tingay 
 \orcidA{}}

\AuthorNames{Steven J. Tingay}

\isAPAStyle{%
       \AuthorCitation{Tingay, S.J.}
         }{%
        \isChicagoStyle{%
        \AuthorCitation{Tingay, S.J.}
        }{
        \AuthorCitation{Tingay, S.J.}
        }
}

\address[1]{International Centre for Radio Astronomy Research, Curtin University, GPO Box U1987, \linebreak  Perth, WA 6845, Australia; s.tingay@curtin.edu.au\\}

\abstract{The Murchison Widefield Array (MWA) is a low frequency radio interferometer designed and developed by an international consortium, operated on behalf of the consortium by Curtin University.  The MWA is a Precursor for the low frequency Square Kilometre Array (SKA) and is located at the SKA site in Western Australia, Inyarrimanha Ilgari Bundara, the CSIRO Murchison Radio-astronomy Observatory.  Commencing science operations in 2013 after an extended development period, the MWA has performed observations over a wide set of science objectives, has been upgraded multiple times, and has played a fundamental role in the development of the low frequency SKA.  As MWA Program Manager from 2008 to 2011, as Director from 2011 until 2015, and then again from 2021 to the present, I describe some personal reflections on the MWA's activities and successes in these different dimensions, as well as my view of some of the approaches that have enabled these successes. I offer some of the lessons I've perceived over the last 17+ years in \linebreak  the project.}

\keyword{Murchison Widefield Array; Square Kilometre Array; radio astronomy \linebreak  instrumentation} 

\begin{document}


\section{Introduction}

The Murchison Widefield Array (MWA\endnote{\url{https://www.mwatelescope.org} (accessed on 10 September 2025)}) is
 one example of a novel new generation of radio telescopes, operating at low radio frequencies (in this case between 70 and 300 MHz, in~the VHF band) and utilising aperture array technologies.  Other examples of such instruments are LOFAR, based in The Netherlands and also distributed across a number of European countries~\cite{2013A&A...556A...2V}, and~the low frequency component of the Square Kilometre Array (SKA;~\cite{5136190}), based in Western Australia.  Aperture array technologies at low radio frequencies provide an extremely efficient method of building large collecting area telescopes at low cost and low complexity for the in-field deployment, because~they utilise simple antennas that have no moving parts.  Aperture arrays, operating as Phased Array Feeds (PAFs) have also been recently utilised to great effect at higher frequencies ($\sim$1 GHz) by telescopes such as ASKAP~\cite{2008ExA....22..151J} and APERTIF~\cite{2022A&A...658A.146V}.

The headline science motivation for building the MWA was the prospect to detect reshifted radio emission from the hyper-fine transition of neutral hydrogen during the so-called Epoch of Reionisation (EoR), potentially shedding light on the conditions in our Universe during the first billion years after the Big Bang~\cite{2006PhR...433..181F}.  As~the MWA project concept and partnership developed, the~science motivations expanded~\cite{2019PASA...36...50B,2013PASA...30...31B}.

The MWA is an official precursor for the much larger planned low frequency SKA, and~the MWA and SKA share the same site, Inyarrimanha Ilgari Bundara, the~CSIRO Murchison Radio-astronomy Observatory, on~the traditional lands of the Wajarri Yamaji in the mid-west of Western Australia.  This location was selected as a site for the SKA in 2012\endnote{\url{https://www.skao.int/en/about-us/91/history-ska-project} (accessed on 10 September 2025)}, based on very low levels of anthropogenic radio frequency interference (RFI), among~other characteristics~\cite{Schilizzi2024}.  Construction of the MWA predated this decision, and~the existence of the MWA project was a positive indicator that the site was good for low frequency radio~astronomy.  

In addition to being connected to the SKA as a Precursor for a large international project, the~MWA itself has always inherently been international, existing as a consortium of member nations and many individual organisations in those nations; in this sense the MWA has been a bottom up development around shared scientific goals.  The~current MWA partner membership is composed of Australia, Canada, China, Japan, Switzerland, and~the USA\endnote{\url{https://www.mwatelescope.org/about/\#partnerships} (accessed on 10 September 2025)}.  Each country contributes financially and scientifically to MWA outcomes.  In~the past, India and New Zealand were members for periods of~time.

The technologies that have underpinned the MWA systems have been comprehensively described in a series of system description papers, describing the initial MWA concept~\cite{2009IEEEP..97.1497L}, the~as-built first phase of the MWA~\cite{2013PASA...30....7T}, and~the upgraded second phase of the MWA~\cite{2018PASA...35...33W}.  A~series of science objectives papers have also been published~\cite{2019PASA...36...50B,2013PASA...30...31B}.  Currently the upgrades for a third phase of the MWA have been completed and an increment on the system description papers will be published in due course (Tingay~et~al., in~preparation).  For~more detail on many of the points described in this review, see these overview papers, as~well as the technical references cited throughout the~review.

The intent of this review is not to re-describe the technical aspects of the MWA in detail, provide a detailed account of the development period, or~collate comprehensive statistics around the MWA's scientific successes, but~selectively describe some of the inflection points in the MWA's journey, some of the approaches that were adopted and decisions that were made by the management teams, and~where that led the program scientifically.  Thus, this is necessarily a rather personal reflection by the author on what has been important {, but~not comparable to a formal completion report, for~example as conducted for the Very Large Array in 1982\endnote{\url{https://library.nrao.edu/public/memos/vla/misc/VLAU\_10.pdf} (accessed on 10 September 2025)}}.  Others who have also made very significant contributions to the MWA's success may select a different set of points to emphasise.  This review is not meant to be definitive or exhaustive in any way, but~largely commentary based on the MWA experiences of the author, which may be useful to inform the planning of future~projects.

\section{Design, Prototyping, Construction, and~Commisioning}

The MWA was originally conceived as being composed of 512 ``tiles'' (the MWA unit of collecting area, 16 dual-polarised bowtie dipoles arranged as a 4 $\times$ 4 grid approximately 5 m on a side, forming an analog beam-steered aperture array), spread over an area contained within approximately a single ionospheric patch~\cite{2009IEEEP..97.1497L}, in~order to simplify calibration for ionospheric effects (approximately 3 km diameter).  Digital receivers, a~Monitor and Control software suite, an~FPGA-based correlator, and~a Real-Time System (RTS) for real-time imaging and calibration completed the main components for the system.  Visibility data were to be discarded following RTS processing, and~no plans for a data archive~existed.

As a project with multiple funding streams through multiple countries and funding agencies, and~a multitude of teams in those countries contributing to design and production efforts, the~complexity and ambition of the original MWA concept came under significant pressure, relative to the available resources, in~2008.  At~that point, a~substantive reshaping of the project scope, budget, schedule, and~risk profile was implemented, following funding agency reviews in the USA and~Australia.

\subsection{Reshaping the~Project}

Some of the key changes, as~I see them, adopted by the management team are briefly described below:

\begin{itemize}

\item A very early decision was made to remove the RTS as a real-time element of the MWA system.  Single pass calibration and imaging had always proved difficult for any interferometric radio telescope.  This was the case in the late 2000s, and~indeed is still the case in 2025.  {In fact, VLA Computer Memorandum 172\endnote{\url{https://library.nrao.edu/public/memos/vla/comp/VLAC\_172.pdf} (accessed on 10 September 2025)} from 1984 provides an unvarnished account of the VLA data processing Pipeline, containing statements that are stunningly relevant to projects underway in 2025.}  A decision was made to conduct the RTS processing on visibilities that had been stored, removing the risk of real-time processing.  The~RTS provided effective MWA data calibration and imaging in this mode for a number of years into the operational period of the final instrument, and~formed the basis for subsequent generations of software for calibration and imaging that were used for both specific experiments (e.g., the EoR experiment) and as a general part of the MWA data processing pipeline~\cite{2017MNRAS.471.3974J}.  {The resolution to the data storage problem created by removal of real-time processing is described in Sections~\ref{Sec:other} and \ref{Sec:asvo}};

\item The MWA sub-system most challenged by a highly distributed workforce was the FPGA-based correlator.  The~{hardware and firmware} developments were complex and featured a large number of interfaces across different teams, none of whom were resourced to be dedicated to this task.  Despite persistence across several years and significant investment, the~development did not converge.  In~the early 2000s, software correlation was re-emerging (some of the earliest digital correlators were implemented in software) as a technology~\cite{2007PASP..119..318D}, and~Graphical Processing Units (GPUs) were starting to make an impact in radio astronomy data processing~\cite{2004ExA....17..287S}.  In~a necessarily bold move, the~FPGA correlator effort was disolved, and~replaced with a GPU-based correlator development.  Within~12 months, a~flexible correlation system had been completed~\cite{2015PASA...32....6O} (informed by prior efforts~\cite{2009PASP..121..857W}).  {Similar facilities around the world were on similar paths at about the same time, for~example LOFAR \citep{2018A&C....23..180B}}. Reverse engineering a software correlator into the MWA system was not trivial, however.  But, as~a by-product, the~MWA Voltage Capture System (VCS) became a possibility, and~proved a boon to science areas such as pulsar research~\cite{2015PASA...32....5T}.  The~GPU correlator and VCS still utilised the fine channelisation hardware produced for the original correlator.  The~MWA has continued to invest in GPU-based correlators, with~an upgrade in 2022 benefiting from several generations of commodity technology improvements~\cite{2023PASA...40...19M};

\item The MWA receiver system was also a design and prototyping exercise between different teams in different countries, again not resourced for dedicated delivery.  However, as~opposed to the correlator development, enough of the receivers had been realised to be reasonably confident that a working system could be completed.  As~project funds came under pressure, an~industry partner was engaged to pull together the disparate receiver elements and undertake design for manufacture for a receiver that could be fielded to the harsh environment of the Western Australia mid-west.  Poseidon Scientific Instruments in Fremantle, Western Australia, undertook this task, and~ultimately delivered working receivers.  They were fielded in 2012 and are largely operating still today, even though they are a long way past their nominal end-of-life~\cite{2015ExA....39...73P}.  The~receivers, however, had design features that could not be changed, the~most notable being that they are critically sampled receivers, meaning that aliasing occurs near coarse band edges (a choice made in the mid 2000s).  This is generally a minor nuisance for most MWA science cases, but~for the EoR experiment the choice has proved to be a significant issue to overcome, as~the flagged data at the band edges represent missing $k_{\parallel}$ Fourier modes of the EoR two-dimensional power spectrum~\cite{10.1093/mnras/stad845}.;

\item The reshaping of the MWA's scope and budget required a balancing of expenditure across the entire system, to~maintain a capable instrument and retain its primary science goals.  The~largest apparent change in scope emerged at the headline level in the number of tiles in the array, and~hence the overall instantaneous sensitivity.  The~costs in question scaled linearly with the number of tiles, in~terms of: the supporting infrastructure; the tiles themselves; and the number of receivers.  And~with the square of the number of tiles for: the correlator; and the subsequent data storage/processing.  Management was faced with a choice to balance these factors.  Ultimately, {management} decided to build a system with 128 tiles (a factor four lower than the originally envisaged 512), but~build the infrastructure to accommodate a doubling of the size of the array to 256 tiles at a later date (see Section~\ref{Sec:phaseII}).;

\item Having settled on 128 tiles, the~next obvious question was how to configure them into a distribution to maximise the science output of the facility.  This decision had serious practical consequences, as~it would drive the cost of the associated infrastructure (trenching for power and fibre reticulation, for~example).  Management consulted the MWA science community for feedback on this question.  With~a reduced amount of collecting area, the~EoR team was a strong advocate for an exclusive concentration of the collecting area on very short baselines, to~sample the angular scales thought relevant for the EoR experiment.  Other science teams, such as the solar science team, advocated for maximising the angular resolution of the array, having long baselines.  {These two ends of the configuration spectrum represented fundamentally different outcomes for the MWA facility: on one hand, an~EoR experiment with few other science cases that could be supported; on the other hand, a~facility with an apparent reduction in sensitivity for the EoR, but~the capability to support many other science cases}.  The~EoR experiment was, and~is, known as a famously difficult experiment.  Thus, management decided to implement a tile configuration with many short baselines but also many long baselines, and~a relatively uniformly filled $(u,v)$ coverage.  Ultimately this decision supported a very broad science case~\cite{2013PASA...30...31B}.  The~EoR team also realised that they needed long baselines to characterise an EoR foreground model, which has proved to be one of the tougher problems facing EoR detection~\cite{2021PASA...38...57L}.  The~final configuration of 128 tiles was determined using techniques developed for the originally envisaged 512 tile array~\cite{2012MNRAS.425.1781B}.  With~a configuration decided, infrastructure costs could be estimated and a final scope, budget, schedule, and~risk profile was developed;

\item One final big change to the project scope is worthy of mention in this section, I believe.  This was not a decision of management, but~was enthusiastically embraced by management.  As~part of the funding agency review in Australia, one of the outcomes was the need to prove to the funding agency (Astronomy Australia Limited) that the MWA team could make a telescope like the MWA work.  Hence, a~gate to final funding was negotiated, the~construction of a one-quarter scale prototype, consisting of 32 tiles (which became known as 32T).  The~requirement was very practically framed, however, in~that it did not require prototype versions of all the final sub-system elements (receivers, correlators, RTS etc), but~analogous versions of these sub-systems that achieved the same function.  32T was progressed over an approximate two year period ($\sim$2009 to $\sim$2011), during~which the 32T system was built using largely simple off-the-shelf analogs of the final envisaged hardware systems (for example, as~described in~\cite{2009PASP..121..857W}).  Most of the sub-systems were much less capable than the final envisaged sub-systems, but~could be built and deployed quickly.  32T had baselines a factor ten less than envisaged, at~a few hundred metres.  32T proved a masterstroke, as~it gave the MWA teams a concrete and near-term physical target that was achievable.  As~it was achieved quickly, 32T collected and distributed real data into the hands of MWA scientists quickly, which built enthusiasm and excitement~\cite{2013ApJ...771..105B,2014MNRAS.438..352B,2010PASP..122.1353O,2012ApJ...755...47W,2011ApJ...728L..27O}.  And, importantly, during~the two years that 32T operated, the~management team had time to consider and execute most of the big decisions mentioned in the bullet points above.  32T achieved real high impact science outcomes, allowed management the space to develop final plans and costs, defined approaches to data calibration and processing, and~nourished the science community that would carry the MWA forward.
\end{itemize}

In addition to the above, many, many other major and minor decisions were made that brought the MWA to the point of being technically ready to enter construction.  The~question of software across the MWA systems warrants an entire description of its own.  The~software suite for the instrument Monitor and Control grew organically, following in most cases hardware development, and~has largely been a bespoke development for the MWA.  Other code bases for detailed technical developments such as the data transport and data archiving have evolved strongly over the project lifetime.  And~other elements of code for functions such as radio frequency interference elimination and flagging, calibration, and~an understanding the primary beam model (among many other aspects) have developed out of the operations team and the wider MWA user community, before~being absorbed into operations.  Software has been the result of an eclectic and organic mix of contributors and iteration, but~anchored by a small and excellent core operations staff, who take ownership and carriage of mission-critical~software.

Omissions in the above by no means diminish relevance, but~are due to a finite page count.  I've tried to pull out major decisions, as~I see them, from~which many other decisions flowed across the system and the contributing~teams.

\subsection{Other~Factors}
\label{Sec:other}

Other influences were important in my view, however, which I {categorise} below as ``other factors''.  By~which I mean that an element of good fortune was involved, rather than simply hard work and prudent decision~making.

\begin{itemize}

\item As noted above, even though the MWA was designed to generate very high data rates, no actionable plans for a data archive existed before approximately 2008.  Indeed, quite a lot of the decision making described above was made in the absence of a data archive plan, which was simply accepted as ``something to solve later'', when and if a telescope could be realised.  As~the MWA project plan was reaching a final state of maturity, the~consequences of the Global Financial Crisis (GFC) of the late 2000s were becoming apparent across the world.  In~Australia, very significant economic stimulus funding was made available in many areas of Commonwealth Government expenditure, including in science research infrastructure.  In~this area of investment, one sub-area was high performance computing, including the establishment in Perth of the Pawsey Supercomputing Centre (Pawsey\endnote{\url{https://pawsey.org.au/} (accessed on 10 September 2025)}).  Pawsey was funded for many reasons, but~one reason was the then upcoming Square Kilometre Array (SKA) and the existence of SKA precursors ASKAP and MWA.  Thus, Pawsey was able to solve immediate data storage and data processing issues for ASKAP and the MWA.  The~MWA was initially allocated large-scale data storage for the MWA data archive, and~then later also secured access to large-scale computing resources.  The~MWA and Pawsey have grown up together, and~have supported mutual success.  At~its peak the MWA data archive consisted of 55 PB of data under curation, made accessible to the world's astronomers via a searchable portal (see Section~\ref{Sec:asvo});

\item In addition to funds for Pawsey as a consequence of the GFC stimulus, funds were also made competitively available for astronomy research infrastructure.  As~the MWA management team had just completed a robust reshaping of the project plan, scope, cost, and~schedule (and successfully completed 32T), we were in a strong position to propose a ``shovel ready'' construction project, which was successful.  These funds, in~addition to funding already secured, allowed the reshaped plan to be realised.  One might say that the project was lucky.  Personally, I don't like the word luck in this context.  To~paraphrase a well-known quote, ``luck favours those prepared to accept it'', so I prefer to describe our situation as fortunate.
\end{itemize}

In addition to the above, quite a lot of creative fund raising was involved.  For~example, the~GPU-based correlator and the VCS were financed with industry support from IBM, but~the support needed to come via IBM New Zealand, which precipitated New Zealand entering the MWA consortium, supporting the growth of a new radio astronomy community in that country\endnote{\url{https://www.techtaffy.com/mwa-consortium-using-ibm-technology-to-explore-universe-origin/} (accessed on 10 September 2025)}.  Bi-lateral Australia - India funding was also secured, to~support elements of the receiver developments\endnote{\url{https://www.industry.gov.au/sites/default/files/2022-08/aisrf\_10\_yr\_anniversary.pdf} (accessed on 10 September 2025)}.

\section{Construction, Commissioning, and~Early~Operations}

32T provided us with time to plan, as~well as critical experience.  Thus, once funding was secured, construction started in late 2011.  Construction reached practical completion by the end of 2012.  This included all infrastructure, deployment of all sub-systems, and~the ability to commence engineering and science commissioning.  This was a stunning achievement on an effective green field site (not green, if~you have ever been to the Western Australian outback) in an extremely remote location.  In~large part, this was due to the very positive contract negotiations we were able to enter into with a mid-west construction contractor, Geraldton Electrical (currently trading as G.Co.), with~whom the MWA still deals in order to support ongoing telescope operations, more than a decade after construction was~completed.

Engineering commissioning was progressed by the teams who designed the sub-systems, as~well as the now-formed MWA operations team.  A~science commissioning team was formed from members of the MWA science community, across different areas of science interest, different organisations, and~different countries.  The~team was highly structured and was provided strict directives.  Again, the~32T experience was invaluable, as~the commissioning team naturally split the 128 tiles up into 4 $\times$ 32 tile sub-arrays and commissioned each (using software, tools, and~insights developed for 32T) separately~\cite{2014PASA...31...45H}, before~commissioning the final and full 128 tile system.  Engineering and science commissioning started near the beginning of 2013 and was complete by approximately the middle of~2013.

Early operations commenced smoothly and scientifically high quality data were produced immediately, appearing in many publications from about 2014 (see publication list URL for the ADS library provided in Section~\ref{Sec:sci}).  The~management team adopted a very conservative approach to releasing MWA observing modes, starting with a small number of modes that would service all science teams, but~reserving the latent flexibility of the MWA until sufficient experience had been gained by the operations team and science community to deal with the additional complexity that comes with~flexibility.

The only significant thing I'll mention about early operations from a management perspective is that at the end of construction, Australian funding switched from construction to operations.  As~such, new funding agreements were required, focused on operations rather than construction.  The~views of operations metrics held at the time were traditional in nature, in~the sense that they were driven in part by the measure of ``hours on sky''.  This is fine for most traditional interferometric radio telescopes (dish-based, higher frequencies, small fields-of-view), but~for the MWA the management team held that this was a poor metric.  With~the MWA's extreme field of view, high sensitivity, and~commensurate very high data rates (in 2012 terms), each MWA observation was capable of supporting multiple commensal science objectives, simultaneously satisfying multiple teams.  MWA management argued that the metric of data volume collected was a better metric, as~that represented the most significant resource bound for the project; we argued for an initial 25\% duty cycle for observations.  After~discussion, this metric and duty cycle was agreed.  Over~time, the~MWA observational duty cycle has increased to well above this initial value and now fluctuates depending on science goals and time of day/year.

The as-built MWA is described fully in~\cite{2013PASA...30....7T} and, recognising the aspiration to more fully populate the established infrastructure with tiles beyond 128, became known as the Phase I~MWA.

\subsection{Phase~II}
\label{Sec:phaseII}

The Phase II aspiration came to pass in due course, as~the MWA consortium pooled financial contributions and leveraged them through a successful competitive funding proposal for research infrastructure in Australia.  This allowed the array to double the number of tiles from 128 to 256, fully populating the infrastructure, but~the additional 128 tiles were also deployed such that the maximum baseline length (and hence angular resolution) was also doubled, from~approximately 3 km to approximately 6~km.  

However, Phase II involved no new receiver systems, so only 128 of the 256 tiles could be operated at any given time.  The~array was therefore physically reconfigured (by physically moving receivers) between a compact configuration of 128 tiles and a long baseline configuration of 128 tiles, approximately every six~months.

The Phase II MWA is comprehensively described in~\cite{2018PASA...35...33W}.

\subsection{The MWA Data~Archive}
\label{Sec:asvo}

Many other significant enhancements were introduced into the MWA system during the Phase I period, very notably to software and Monitor and Control systems.  One fundamental enhancement was the introduction of the MWA All-Sky Virtual Observatory (ASVO;~\cite{2022arXiv220310710O}) portal\endnote{\url{https://asvo.mwatelescope.org/} (accessed on 10 September 2025)}, as~the front end to the growing MWA data archive resident at Pawsey.  The~ASVO development, and~the continued improvement of automated data calibration and RFI flagging algorithms, meant that over the Phase I and II MWA operations periods, data were more easily discoverable, accessible, and~of higher quality.  Quite a lot of the reason for this was a very positive feedback loop between the MWA science teams and the MWA operations team.  As~science users developed and refined data processing pipelines, knowledge from that experience was fed into operations, where it was incorporated into the MWA's standard systems, for~the benefit of all users.  The~ASVO is very mature and heavily utilised at this point in time, and~has supported the MWA very capably, as~the archive reached 55 PB in 2025.  The~MWA ASVO node succeeded an earlier archive that provided initial service to the MWA~\cite{2013ExA....36..679W}.

\section{Science~Implications}
\label{Sec:sci}

The choice made to implement the MWA as more than a narrowly focused EoR experiment has paid dividends (including to the EoR project).  The~long baselines of the MWA in Phase I and Phase II have supported large-scale wide-band continuum surveys, in~the form of GLEAM~\cite{2019PASA...36...47H,2017MNRAS.464.1146H, 2015PASA...32....5T} and GLEAM-X~\cite{2022PASA...39...35H, 2024PASA...41...54R}, from~which a great many science projects have been realised.  Solar studies (for example~\cite{2018NatSR...8.1676C}) and studies of transient radio sources (for example~\cite{2022Natur.601..526H, 2023Natur.619..487H}) have prospered from use of the long baselines, wide fields of view, and~high survey speed at low frequencies. {I gloss over so many iterations of learning when it comes to the myriad challenges overcome in understanding the instrument, including calibration.  But, certainly, conducting all-sky continuum surveys proved an excellent way to uncover and resolve these many, often subtle, issues.}

Despite the critically sampled receiver system, the~EoR project with the MWA is internationally competitive for this extremely difficult project.  The~MWA EoR team has made fundamental contributions to the international pursuit of the first EoR detection, publishing a large number of the MWA's most highly cited publications (for example~\mbox{\cite{2020MNRAS.493.4711T, 2016ApJ...833..102B, 2014PhRvD..89b3002D}}.  Also utilising the MWA's compact Phase II configuration have been pulsar studies (for example~\cite{2017ApJ...851...20M}).

The MWA's areas of science contribution are too many to list in this review.  Overviews of a subset of Phase I and Phase II science ambitions and results are provided in detail in~\cite{2013PASA...30...31B} and~\cite{2019PASA...36...47H}.  A~publicly accessible library of MWA papers is available via ADS 
 (\url{https://ui.adsabs.harvard.edu/public-libraries/SB5\_iHeZTxCZDCj5kUuR6Q} (accessed on 10 September 2025)), which is regularly updated.  In~summary, as~of 8 July 2025, the~library contained 390 publications (373 produced since the 2013 start of operations), with~an H-index of 57 and a total citation count of 18,235.  The~top ten cited papers range from 250 citations to 3639 citations and span MWA system description papers, the~GLEAM survey, Fast Radio Burst follow up, Gravitational Wave event follow up, MWA science definition papers, and~software~developments.

\section{Supporting the Square Kilometre~Array}

The MWA, as~a Precursor for the SKA, has fulfilled a number of roles.  The~concept of the SKA Precursor emerged as a number of facilities around the world developed SKA technology development programs, with~those facilities gaining ``Pathfinder'' status; selected Pathfinders located at the two SKA sites in Western Australia and South Africa were given Precursor~status.

As an SKA Precursor, in~particular as a Precursor for the low frequency component of the SKA, the~MWA has taken significant strides toward key areas of SKA science, in~particular in low frequency studies of pulsars, transients, the~EoR, and~in low frequency surveys, as~described above.  The~MWA has deeply studied the ionosphere at the SKA site, and~has developed methods for characterising the ionosphere and mitigating its effects on radio astronomy data~\cite{2015GeoRL..42.3707L, 2017MNRAS.471.3974J}.  Through the {pursuit} of leading science, the~MWA has fostered the growth of the low frequency community within Australia and all other MWA partner countries.  This community helps lay the base for the SKA-Low community of the~future.

The MWA has been the platform for training and up-skilling a new generation of engineers, operations scientists/technicians, and~researchers.  These individuals are now leading members of the SKA community and, indeed, as~the SKA Observatory (SKAO) has dramatically ramped up hiring to support construction, commissioning, and~science operations, many of these individuals have won those~roles.

The MWA's lean operations model has, by~necessity, meant that the project has relied heavily on key partnerships with industry and commercial providers, over~a long period of time.  So, the~up-skilling effect has been in the commercial sector as well; these companies, including businesses local to the SKA in Western Australia, have now won significant SKA~contracts.

Fundamentally, the~MWA infrastructure established to support the telescope was used over a period of more than a decade, to~host multiple generations of the Aperture Array Verification System (AAVS) hardware, effectively full SKA-Low station prototypes~\cite{2013ursi.confE...1H, 2015ITAP...63.5433S, 2017PASA...34...34W, 2021A&A...655A...5B, 2022JATIS...8a1010W, 2022JATIS...8a1014M}.  The~AAVS program started as an SKA design activity, which progressed to substantial SKA pre-construction contracts that delivered tender ready design packages to be let by the SKAO to industry.  This was an inherently international activity, involving a range of SKA partner countries, but~the deepest involvement was from Australia and Italy.  The~SKA-Low stations would not be ready for SKA construction without the AAVS program, which would not have been possible without the MWA.  Ernst and Young estimated an approximate \$AU35M cost saving to the SKA from this program, due to the advances in Technical Readiness Level (TRL) from the AAVS developments\endnote{\url{https://mwatelescope.org/wp-content/uploads/2022/08/Curtin\_University\_-\_The\_economic\_and\_social\_impact\_of\_the\_MWA.pdf} (accessed on 10 September 2025)}.

Finally, MWA data products from GLEAM, and~the wider experience brought by people trained on the MWA, made significant contributions to the realisation of the first SKA-Low image, released in June 2025\endnote{\url{https://www.skao.int/en/news/621/ska-low-first-glimpse-universe} (accessed on 10 September 2025)}, (2025, George Heald, personal communication).

As the SKA-Low progresses in its construction and commissioning phase, and~into early science in the late 2020s, the~MWA is well placed to continue to support the goals of the SKAO, via both accumulated experience and the science the MWA is~generating.

\section{Lessons Verified/Learned (by the Author)}

It is 12 years since MWA operations started in 2013, and~17 years since the re-shaping of the MWA project plan commenced in 2008.  It is interesting to reflect on the lessons learned, big and small.  Early on, Tom Booler recorded some lessons learned from the MWA planning and construction phase\endnote{{\url{https://github.com/steven-tingay/Booler-Lessons-MWA/blob/main/Booler-Lessons-MWA-small.pdf} (accessed on 10 September 2025)}}, all of which still hold today, I~believe.

At a macro level, from~my perspective, the~MWA ultimately prospered as a result of some of the challenges that the project faced, and~the reactions to those challenges.  For~example, by~way of necessity and tight budgets, the~project was effectively forced to seek solutions from a wide range of sources, from~industry, from~a small number of individuals with a particularly deep and broad technical knowledge and a practical delivery mindset, and~from the wider MWA~community.

Each of these dimensions proved critical to the MWA's success, in~my view.  The~small number of individuals that made up the operations team and management team following the re-shaping had a large amount of practical experience in project delivery, and~a deep knowledge of radio astronomy science and technology; not necessarily all in the same individual, but~across the small team.  This allowed a tight focus and short lines of communication/coordination, which reduced decision making times and allowed rapid trade-offs to be~examined.  

The involvement and input of the wider MWA science community was also critical{, and~this was recognised by the management team almost immediately.}  Ultimately the management team was delivering an instrument for the community, and~the community needed to provide advice on what was delivered.  The~constant challenge for management was in balancing community needs (and desires) with practical constraints.  I think through clear communication both ways with an understanding community, and~being open to accepting and seriously considering input, the~MWA reached a good place.  This communication engaged the community, such that they were enthusiastic to work on MWA data and science.  They felt connected to the instrument, through inclusive exercises such as the science commissioning process.  The~MWA needed an engaged and connected community in order to be scientifically productive.  And~at various points we needed the community to help with technical developments for operations, so they needed to be technically competent with the full MWA~system.

With a small but expert team, and~limited long-term financial commitments in place, the~MWA project in some ways was forced to outsource development tasks to industry.  This suited the mindset of the management team - why have astronomers do things that industry is better at providing?  But this situation also suited the project constraints well.  For~example, in~order to deliver the receivers via Poseidon Scientific Instruments, we entered into a sequence of small development contracts before issuing a big production contract.  We structured small, incremental investments until both parties had confidence that delivery was possible.  This allowed us to be agile, plan our finances, and~live within the constraints we faced.  Over~time, our work with industry partners has been integrated into our long-term operations strategy, providing us with an extra-project workforce, expert in our site and our instrument, but~not full time on our salary~books.

If I had to sum up the above, I think the main lessons that were verified for me were that a project needs to think hard about who is best placed to deliver, and~be prepared to act on the conclusions of that thinking.  Only use internal resources for important aspects of design and requirements specification, outsource to industry where possible, adopt commercial off-the-shelf standard technologies where possible (don't reinvent the wheel yet again), and~only have internal delivery of hardware where there is no alternative.  Conversely, when it comes to software, I've learned that this seems to be more difficult to specify and outsource; in my experience practical decisions need to be made about hardware, and~software development tends to follow these choices.  Recognising this, and~building this approach into the MWA program, avoided over-investment too early in software, which always runs the risk of being wasted if the hardware is not settled.  Software considerations always, however, need to be taken into account in hardware decisions.  Software is, therefore, very important to have embedded within the project, in~order to time investments in software development optimally.  Some of this resonates with project analysis as far back as the early 1980s\endnote{\url{https://www.atnf.csiro.au/observers/memos/d93388~1.pdf} (accessed on 10 September 2025)}.

I think I've also come to understand that the radio astronomy community, particularly the interferometry community, is a bit different to communities at some other wavelengths.  They really want to be at the coalface of the system, with~their hands on the data and data processing, with~the ability to modify processing pipelines and refine how the data are collected and processed.  This community is not really happy to just accept a data product in ``science ready'' form, as~is the case in many other~communities.

I find this to be a very interesting observation I've noted over the years.  In~the case of the MWA, by~necessity the project had to embrace this characteristic of our community.  We are now infinitely better for it, in~my view.  The~lines between operations team and science teams are quite blurry in some dimensions, with~innovations and knowledge flowing freely both ways, while it is clear who ultimately controls the~system.

I put this apparent community characteristic down to a few things.  The~development of interferometry was a real hands-on experimental endeavour, where you needed to understand all aspects of the instrument in order to do science.  Radio interferometry is still a relatively young field, so many of the pioneering individuals are still active in the community and continue to keep this part of interferometry's original culture alive.  It has certainly heavily influenced me and others of my generation.  So, I think some of the community attitude stems from early~influences.

With the boon in new, and~increasingly powerful, interferometers over the last 25 years around the world, including in new centres for radio astronomy such as South Africa, a~very large number of individuals have become extraordinarily proficient in advanced interferometric techniques.  As~interferometers are extremely flexible, so some of this knowledge is extremely specialised.  The~majority of interferometry expertise lies outside formal telescope operations~teams.

I've come to believe that these characteristics of our community are a valuable and quite unique asset and should be maximally leveraged as part of planning for future projects.  That is not to say the leverage is without~challenges.

\section{Phase~III}

Lest the reader feel that the introspection in this paper signals that the MWA is at the end, as~the title of the paper indicates the MWA is just maturing.  The~MWA has just completed a Phase III upgrade, after~a five year design, prototyping, and~production period.  Phase III is centred around a new correlator~\cite{2023PASA...40...19M}, as~well as a new receiver suite that will mean that all 256 tiles will be able to be utilised simultaneously.  The~new receivers are oversampled and will service the short baselines of the MWA, to~resolve the challenges mentioned above when it comes to the EoR experiment.  A~full Phase III system description paper will appear in due course.  This review is not intended to pre-empt that description, but~the Phase III period is currently underway and it is worth noting here that the experience and lessons of Phases I and II heavily informed how Phase III was successfully~delivered.

\vspace{6pt}

\funding{The work to prepare this manuscript received no external funding}

\acknowledgments{This scientific work uses data obtained from Inyarrimanha Ilgari Bundara/the Murchison Radio-astronomy Observatory. We acknowledge the Wajarri Yamaji People as the Traditional Owners and native title holders of the Observatory site. Establishment of CSIRO’s Murchison Radio-astronomy Observatory is an initiative of the Australian Government, with~support from the Government of Western Australia and the Science and Industry Endowment Fund. Support for the operation of the MWA is provided by the Australian Government (NCRIS), under~a contract to Curtin University administered by Astronomy Australia Limited. This work was supported by resources provided by the Pawsey Supercomputing Research Centre with funding from the Australian Government and the Government of Western Australia. The~International Centre for Radio Astronomy Research (ICRAR) is a Joint Venture of Curtin University and The University of Western Australia, funded by the Western Australian State government}

\begin{adjustwidth}{-\extralength}{0cm}
\printendnotes[custom]

\reftitle{References}

\PublishersNote{}
\end{adjustwidth}
\end{document}